\documentclass[epj,final]{svjour}

\usepackage{latexsym}
\usepackage{url}
\usepackage{amsfonts}

\RequirePackage{graphicx}

\begin{document}
\title{A principal possibility for computer investigation of evolution of dynamical systems with independent on time accuracy}
\author{V. G. Gurzadyan\inst{1,2}, V.V.Harutyunyan\inst{2}, A.A. Kocharyan\inst{2,3}
}                     
%
%
\institute{SIA, Sapienza University of Rome, Rome, Italy \and Center for Cosmology and Astrophysics, Alikhanian National Laboratory and Yerevan State University, Yerevan, Armenia \and 
School of Mathematical Sciences, Monash University, Clayton, Australia}
\date{Received: date / Revised version: date}
%

\abstract{
Extensive N-body simulations are among the key means for the study of numerous astrophysical and cosmological phenomena, so various schemes are developed for possibly higher accuracy computations. We demonstrate the principal possibility for revealing the evolution of a perturbed Hamiltonian system with an accuracy independent on time. The method is based on the Laplace transform and the derivation and analytical solution of an evolution equation in the phase space for the resolvent and using computer algebra. 
} 

%
\maketitle
\section{Introduction}

The appearance of powerful computers has made the N-body simulations among efficient tools \cite{A} for the study of various astrophysical and cosmological problems.  The N-body gravitating systems cover a broad class of astrophysical problems, from the evolution of the Solar system as of a nearly integrable problem, up to the dynamics of star clusters, galaxies, and galaxy clusters as non-integrable problems. The cosmological simulations include from the evolution of primordial density perturbations up to the formation of dark matter galaxy halos, etc. 

The N-body simulation activities have been developed in two conventional directions:

1.	Numerical integration of N equations of motion in order to follow the evolution of the system. The main difficulty
of these investigations based on standard iterative methods of numerical integration of differential equations (Runge-Kutta, etc.) is the inevitable storage of errors with the increasing of the number of steps. This leads to rapid loss of the reliability of the results derived,  the faster, the larger is the number of equations (i.e. the number of particles) and/or the longer is the duration of the calculations.

2.	Investigation of the character of motion in the systems, which includes the finding out of integrals of motion, analysis of
appearing stochastic regions, etc. The main feature of this direction is the application of methods of dynamical systems,
e.g. of Kolmogorov-Arnold-Moser (KAM) theory, Lyapunov exponents, KS-entropy, etc \cite{Arn}. N-body gravitating systems, for example, do reveal variety of complex, chaotic properties which play a crucial role in their evolution and structure; see e.g. \cite{GS,GK,GK1,GK2,L,L1}.

Speaking of numerical investigations of dynamical systems, both, smooth and discrete, one may note that the second direction seems to be particularly fruitful. Beginning from the pioneer study by Fermi, Pasta and Ulam in the early 1950s, it has led to unexpected and profound results such as the strange attractors, the Feigenbaum sequence of bifurcations, etc. As compared to those remarkable results, the numerical (iterative) study of the evolution of Hamiltonian systems, due to the reasons mentioned, looks moderately successful and reliable. The latter is equally true for such an important class of Hamiltonian systems as the so-called nearly integrable systems \cite{Arn}
\begin{equation}
H(I,\vartheta,\beta) = H_0(I)+\beta H_1(I,\vartheta)\,,      
\end{equation}
where $I, \vartheta$ are the action-angle coordinates and $\beta$ is the small parameter defining the non-integrable part of the Hamiltonian $H_1$. 

The well-known Henon-Heiles system \cite{HH} modeling the stellar motion in a spiral galaxy is of this type; the same class of Hamiltonians can be attributed also to various aspects of planetary dynamics.
The unique role of the system (1.1) in the Hamiltonian mechanics is determined undoubtedly by the proof of the KAM theorem, according to which, when distinct conditions are satisfied, there exist invariant tori,
\begin{eqnarray}
I= I_0 + u(\vartheta, \beta)\, ,  \\
\vartheta = \vartheta_0 + \omega t + v(\vartheta,\beta)\, ,
\end{eqnarray}
where $u,v$  are periodic functions \cite{Arn}.

It is well known that the KAM theorem with its roots goes back to the main problem of the celestial mechanics, the stability and evolution of the Solar system. Although the KAM theorem does not directly reveal the fate of the Solar system, its ideas were used for efficient numerical studies of the Solar system dynamics; see \cite{L,L1} and other studies by Laskar et al. The KAM theorem itself does not determine the value of the perturbation $\beta$, for which its statements are true, and regarding the planetary dynamics there are also principal difficulties in checking of the necessary conditions of the theorem with respect to each consideration. Moreover, due to the Arnold diffusion - a universal instability peculiar to non-linear systems of dimension $N > 2$ - irrespective of whether the Solar system satisfies the conditions of the KAM theorem or not, it still cannot remain stable. Finding itself after an accidental perturbation in the stochastic region of phase space, the system can remain in it for an infinitely long time and therefore, the observed picture has to be destroyed anyway - the planets must either fall onto the Sun or fly away. So, those fundamental results at least do predict that, strictly speaking, the Solar system cannot last forever anyway, although the time scale of the latter instability (Arnold diffusion), according to estimates, far exceeds the Hubble time.

The example described above shows, on the one hand, the universality of systems with a perturbed Hamiltonian; on the other hand, the difficulties in direct application of the KAM theorem for revealing of the evolution of real physical and astrophysical systems.

Connecting certain hopes with computers and speaking of their potentialities, one must mention the importance of computer algebra methods, which, in our opinion, offer new prospects for the investigation of complex dynamics.

Below, we will show that computer algebraic methods along with those of dynamical systems enable the principal possibility of investigation of the evolution of a system without the effect of the storage of errors, i.e. of an accuracy independent on time. The search of error-free numerical integration schemes can be considered among the key problems of stellar dynamics, Problem 5 in \cite{G}; see also \cite{Ein}.

\section{Evolution in the phase space}

The convenience of the investigation of a Hamiltonian system with small perturbation for action-angle variables is connected with the following. As is well known, the Hamiltonian of an n-dimensional integrable system with a phase space $R^{2n} =\{p,q\}$ , e.g. in the case of free oscillators with perturbation
\begin{equation}
H(p,q) = \sum^n_{k=1}\left[\frac{p_k^2}2 + \frac{(\omega^k q^k)^2}2\right] +\beta U(q)\, ,
\end{equation}
i.e. a system having $n$ first integrals of motion in involution, enables one to transit from the variables $(p,q)$ to that of action-angle variables $(I,\vartheta)$
\begin{equation}
\begin{array}{l}
q^k=(2I_k/\omega^k)^{1/2} \cos \vartheta^k\, ,\\[5pt]
p_k=(2I_k\omega^k)^{1/2} \sin \vartheta^k\,.
\end{array}
\end{equation}
Then, for $\beta = 0$ the equations of motion
\begin{equation}
\begin{array}{l}
\dot{I}_k= -\frac{\partial H}{\partial\vartheta^k}\, ,\\[5pt]
\dot{\vartheta}^k= \omega^k(I) + \frac{\partial H}{\partial I_k}\, ,
\end{array}
\end{equation}
have a trivial solution,	
\begin{equation}
\begin{array}{l}
I_k = const\, ,\\[5pt]
\vartheta^k = \omega^k + \delta^k,\, \delta^k = const\, .	
\end{array}
\end{equation}

As it follows from the Liouville theorem, in this case the phase space of the system splits into compact manifolds diffeomorphic to n-tori
\begin{equation}
Tor^n=\{\vartheta^i\; mod\; 2\pi\}.      
\end{equation}

When $\beta\neq0$, the Hamiltonian equations read
\begin{equation}
\begin{array}{l}
\dot{I}^k= -\frac{\partial H(I,\vartheta,\beta)}{\partial \vartheta^k}\, ,\\[5pt]
\dot{\vartheta}^k= \frac{\partial H(I,\vartheta,\beta)}{\partial I_k}\, ,
\end{array}
\end{equation}
where
\begin{equation}
H(I,\vartheta,\beta) = \sum^n_{k=1} \omega^k I_k + \beta V(I,\vartheta)\, .
\end{equation}

However, instead of solving or analyzing these equations as is done conventionally, we will proceed as follows.  

One can observe that if (see \cite{Arn}, Section 39)
\begin{equation}
\dot{x}^k=A^k(x),
\end{equation}
then for any smooth function $f(x)$ the following holds:
\begin{equation}
\frac{d}{dt} f(x)=\frac{\partial f(x)}{\partial x^k}\dot{x}^k=\frac{\partial f(x)}{\partial x^k}A^k(x)\ .
\end{equation}
Therefore,
\begin{equation}
i\frac{d}{dt} f(x) -iA^k(x)\frac{\partial f(x)}{\partial x^k} = 0\ 
\end{equation}
or
\begin{equation}
i\frac{d}{dt} f(x) + L_A f(x) = 0
\end{equation}

where
\begin{equation}
L_A =  -iA^k(x)\frac{\partial}{\partial x^k}\ .
\end{equation}

Considering the function $f(I,\vartheta)$  in the phase space $(I,\vartheta)$ and applying the above observation we find that the equation describing the evolution of this function at $(I,\vartheta)\equiv x$ changing according to (6), has the form 
\begin{equation}
i\frac{d}{dt} f(x,t) + L(\beta) f(x,t) = 0,     
\end{equation}
where
\begin{equation}
\begin{array}{l}
L(\beta) = L_0 + \beta B\, ,\\[5pt]
L_0 = -i\; \frac{\partial H}{\partial I_k}\; \frac{\partial}{\partial \vartheta^k}\, ,\\[5pt]
B(V) = -i \left[\frac{\partial V}{\partial I_k}\; \frac{\partial}{\partial \vartheta^k} - 
\frac{\partial V}{\partial \vartheta^k}\; \frac{\partial}{\partial I_k}\right]\, .
\end{array}
\end{equation}

By means of the Laplace transform 
\begin{equation}
\tilde{f}(x,\lambda) = \int_0^{\infty} dt\, f(x,t)\, e^{i\lambda t}, 
\end{equation}
one can rewrite equation (16) as follows:
\begin{equation}
 -\lambda\,\tilde{f}(x,\lambda) + L\,\tilde{f}(x,\lambda) = -i\, f(x,0)
\end{equation}
or
\begin{equation}
\begin{array}{l}
\tilde{f}(x,\lambda) = i R_\lambda(\beta) g(x)\, ,\\[5pt]
g(x) \equiv f(x,0)\, , 
\end{array}
\end{equation}
where the resolvent $R_{\lambda}(\beta)$ is equal to
\begin{equation}
R_{\lambda}(\beta) = \left(\lambda - L(\beta)\right)^{-1}\, .
\end{equation}

One can show that the following expression is fulfilled
\begin{equation}
\begin{array}{l}  
R_\lambda(\beta)g(x) = \left(\lambda-L_0-\beta B\right)^{-1}g(x)\\[5pt]
=\left(1-\beta R_\lambda B\right)^{-1} R_\lambda g(x)\\[5pt]
=\sum^N_{k=0} a_k(\beta^k)[R_{\lambda}B]^k R_{\lambda}g(x)+o(\beta^N)\, ,
\end{array}
\end{equation}
i.e. at fixed $\beta$, by variation of $N$, we can reach a given accuracy.

If
\begin{equation} 
g(I, \vartheta) = \sum_{k \in Z^n} g_k(I)\, e^{i <k,\vartheta>}\, ,
\end{equation}
where
$$
<k,\vartheta>=k_i \vartheta^i\, ,
$$
then
\begin{equation}
\begin{array}{l}  
R_{\lambda} g(I,\vartheta) = \left(\lambda-L_0\right)^{-1}g(I,\vartheta)\\[5pt]
= \left(\lambda+i <\omega,\partial_\theta>\right)^{-1}g(I,\vartheta)\\[5pt]
= \sum_{k \in Z^n} g_k(I)\left(\lambda+i <\omega,\partial_\theta>\right)^{-1} e^{i <k,\vartheta>}\\[5pt]
= \sum_{k \in Z^n} g_k(I) \frac{e^{i <k,\vartheta>}}{\lambda - <k,\omega>},
\end{array}
\end{equation}
i.e. (18) can he calculated relatively easily. 

We choose the function $g(x)$ in the form
\begin{equation}
g(I,\vartheta)=-\frac{1}{2}\sum_{k=1}^n(I_k-I_{0k})^2+\sum_{k=1}^n\cos(\vartheta^k-\vartheta^k_0)\, .
\end{equation}
This choice is conditioned by two factors: first, it has a single maximum on $R^n \times Tor^n$ and hence, its evolution has the same property; second, it allows an elementary representation in the form of a Fourier expansion.

Having $R_{\lambda} g$, by means of the inverse Laplace transform we get the desired function,
\begin{equation}
f(x,t)=\int_C\frac{d\lambda}{2\pi}\, e^{-i\lambda t}\tilde{f}(x,\lambda)=
 - \int_C \frac{d\lambda}{2\pi} e^{-\lambda t} R_{\lambda}(\beta) g(x)\, ,
\end{equation}
where the contour of integration is
\begin{equation}
C=\{ \sigma +is,\, -\infty < \sigma < \infty,\, s= a>0\}\, .
\end{equation}
Therefore, the evolution of the initial function of $f(x,0) = g(x)$, is found. Estimating after this the maximum of the function
$f(x,t) \equiv f(I,\vartheta, t)$
for each $t$ we find the time evolution of $I,\vartheta$, thus determining the evolution of the Hamiltonian system for those time instances. 

The principal difference between this and the standard iteration methods of integration of Hamiltonian systems is clear; here the time $t$ is a parameter of $f(x,t)$   and does not influence the accuracy of calculation of the latter.
 
\section{Computations via the new scheme}

Firstly, let us note the importance of application of the computer algebra in our calculations. Due to the absence of errors at an integration and differentiation operation, the final results practically do not contain any errors.
To demonstrate this, we consider a dynamical system given by a two-dimensional Hamiltonian of two oscillators with small perturbations,
\begin{equation}
H = \omega_1 I_1 + \omega_2 I_2 +\beta I_1 \cos \vartheta_1 .
\end{equation}

The initial function $f(I, \vartheta, 0)$  was chosen in the form of (25). Then the evolution of that function by means of the procedure described was obtained,
\begin{equation}
\begin{array}{l}
f(I,\vartheta,t)=\exp(i\vartheta_1)(I_1-I_1(0))I_1+\frac{1}2\exp(-i\vartheta_1(0))
[\exp(-i\omega_1t)-1]\\[5pt]
+\exp(i\vartheta_1)(I_1-I_1(0))I_1+\frac{1}2\exp(3i\vartheta_1-2i\omega_1t-i\vartheta_1(0))\times\\[5pt]
\times[\exp(i\omega_1t)-1]\,\omega_1^{-1}\beta-\frac{1}{2}(I_1-I_1(0))^2-\frac{1}{2}(I_2-I_2(0))^2\\[5pt]
+\frac{1}2\exp(i\vartheta_1-i\omega_1t-i\vartheta_1(0))+\frac{1}{2}\exp(i\vartheta_2-
i\omega_2t-i\vartheta_2(0))\, . 
\end{array}
\end{equation}

The evolution of the surface determined by the function $f(I, \vartheta, t)$  at different instances of time and for $I_1=const, I_2=const$ is shown in Fig. 1. 
In Figs. 2 and 3 the variations of  $\vartheta_1$ and $I_1$ by the time for different initial conditions and values of $\beta$  are shown. The trajectories are regular with $2\pi$ period.
During computations, at the increase (decrease) of  $\beta$  by an order of magnitude,  the calculation time varied proportional, at the accuracy $10^{-4}$. The performed computations were little time consuming (for i7, 2600 3.4 GHz processor of 6 GB memory); we plan to apply the approach to real physical systems and provide more results in forthcoming papers. 

Let us stress again that the accuracy of the computations of the trajectory does not depend on time, since no iterations are involved.

\begin{figure}[hbt]
\center{\includegraphics[width=0.8\linewidth]{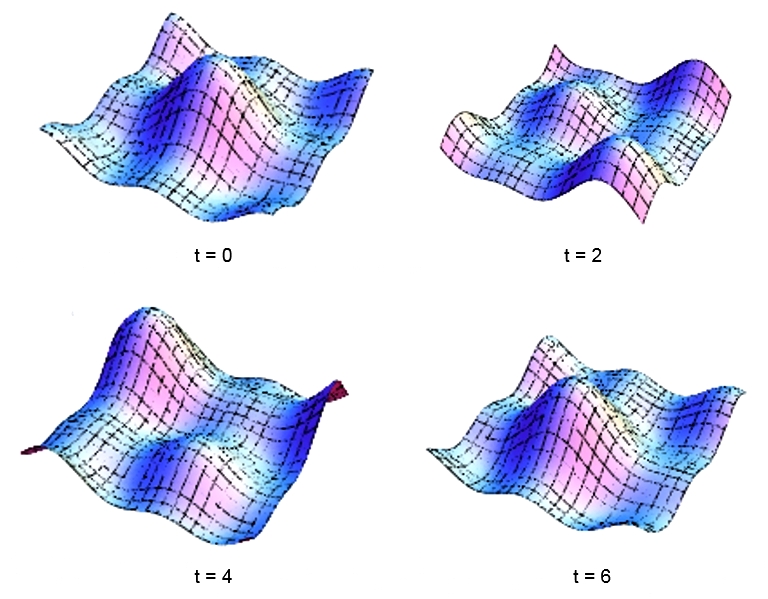}}
\caption{The time evolution of the surface determined by Eq.(29) for $I_1=const, I_2=const$.}
\end{figure}

\begin{figure}[hbt]
\center{\includegraphics[width=0.6\linewidth]{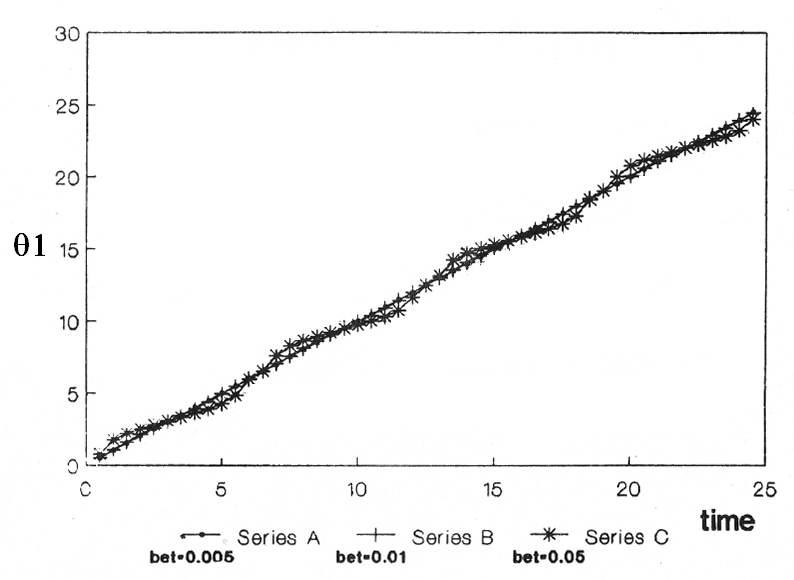}}
\caption{The time variations of $\vartheta_1$ for initial conditions $I_1(0)=1.0, I_2(0)=2.0$ and $\vartheta_1(0)=3.0, \vartheta_2(0)=2.0$ and several values of $\beta$.}%
\end{figure}

\begin{figure}[hbt]
\center{\includegraphics[width=0.6\linewidth]{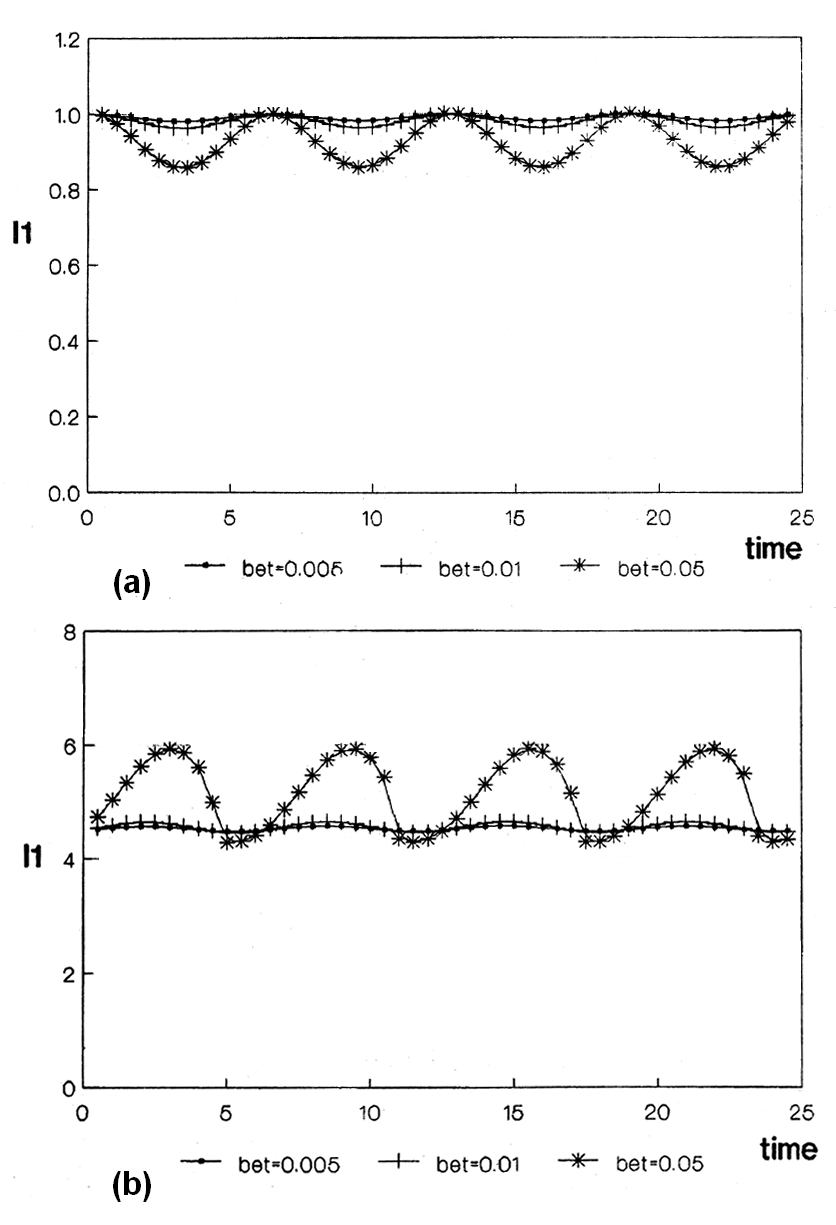}}
\caption{The time variations of $I_1$ for different initial conditions and values of $\beta$: (a)  $I_1(0)=1.0, I_2(0)=2.0$ and $\vartheta_1(0)=3.0, \vartheta_2(0)=2.0$; (b) $I_1(0)=4.5, I_2(0)=2.0$ and $\vartheta_1(0)=1.0, \vartheta_2(0)=2.0$}
\end{figure}

\section{Conclusions}

We demonstrated a principal way of finding out of the evolution of a dynamical system independent on time accuracy; namely, the phase space point  $I_1(0), I_2(0), \vartheta_1(0), \vartheta_2(0)$ describing the initial state of the Hamiltonian system is transferred to an arbitrary $t$, without any storage of errors. The key feature of this method is that we transfer the function $f(I, \vartheta, 0)$ and, therefore, solve an equation which is sufficiently easier (linear), as compared to the Hamiltonian equations. Moreover, the former equation is solved analytically and not by an iteration procedure.

Another remarkable advantage of this procedure is the following: the initial state of a non-linear system during evolution can find itself in the stochastic region of phase space. The description of this is impossible by integration (even numerical) of the Hamiltonian equations, while our method allows us to investigate even this phenomenon. This aspect is particularly important for the study of N-body gravitating systems in view of their well-known chaotic properties. Then this method can open new principal possibilities for cosmological and extensive astrophysical N-body simulations.


\begin{thebibliography}{00}



\bibitem{A} S.J. Aarseth,  Gravitational N-Body Simulations, Cambridge University Press, Cambridge (2010) 

\bibitem{Arn} V.I. Arnold, Mathematical Methods of Classical Mechanics, Springer, New York (1989)

\bibitem{GS} V.G. Gurzadyan, G.K. Savvidy,  A \& A 160, 203 (1986) 

\bibitem{GK} V.G. Gurzadyan, A.A. Kocharyan,  Astrophys. \& Space Sci. 135, 307 (1987) 

\bibitem{GK1} V.G. Gurzadyan, A.A. Kocharyan, A \& A 205, 93 (1988) 

\bibitem{GK2} V.G. Gurzadyan, A.A. Kocharyan, A \& A 505, 625 (2009)

\bibitem{L} J. Laskar, Nature 338, 237 (1989) 

\bibitem{L1} J. Laskar, A \& A 287, L9 (1994) 

\bibitem{HH} M. Henon, C. Heiles,  Astron. J., 69, 73 (1964)  

\bibitem{G} V.G. Gurzadyan, 10 Key problems, in: "Ergodic Concepts in Stellar Dynamics", Eds. V.G. Gurzadyan, D. Pfenniger, Lecture Notes in Physics, vol. 430 (Springer, New York, 1994) 281;  arXiv:1407.0398 

\bibitem{Ein} M. Eingorn, Adv. High Energy Phys., 903642 (2014); arXiv:1409.0220











\end{thebibliography}
\end{document}